\documentclass[fleqn,12pt,twoside]{article}
\usepackage[headings]{espcrc1}
\usepackage{graphicx}
\usepackage{amssymb}
\usepackage{epsfig}

\newcommand{\beq}{\begin{equation}}
\newcommand{\eeq}{\end{equation}}
\newcommand{\bea}{\begin{eqnarray}}
\newcommand{\eea}{\end{eqnarray}}

\newcommand{\bx}{{\vec x}}

\newcommand{\AmS}{{\protect\the\textfont2
  A\kern-.1667em\lower.5ex\hbox{M}\kern-.125emS}}

\title{Progress in nonequilibrium quantum field theory III}

\author{J\"urgen Berges\address{
Institute for Nuclear Physics\\
Darmstadt University of Technology\\
Schlossgartenstr. 9, 64289 Darmstadt, Germany}
and
Szabolcs Bors\'anyi\address{Institute for Theoretical Physics\\
University of Heidelberg,\\
Philosophenweg 16, 69120 Heidelberg, Germany}
}

\runtitle{Progress in nonequilibrium quantum field theory III}
\runauthor{Berges and Bors\'anyi}
\begin{document}
\maketitle 
\begin{abstract} 
We review recent developments and open questions for the
description of nonequilibrium quantum fields, continuing
hep-ph/0302210 and hep-ph/0410330 \cite{sewm0204}. 
\end{abstract} 

\section{Introduction and overview}

Nonequilibrium quantum field theory is a research area showing
substantially increased theoretical activity in recent years.
High-energy particle physicists as well as cosmologists are
working on very similar techniques and sometimes even on the same
underlying nonequilibrium phenomena. This extends to other rapidly
evolving topics such as nonequilibrium dynamics in laboratory
experiments of ultracold quantum gases~\cite{Gasenzersewm}.

Many developments are triggered by high-energy physics
related to collision experiments of heavy nuclei (``Little Bang'')
and early universe cosmology (``Big Bang''). Despite substantial
differences, there are some remarkable parallels. An important
example concerns the role of nonequilibrium instabilities for the
process of thermalization. Though the
origin of these instabilities can be very different, they always
lead to substantial growth of occupation numbers in long
wavelength modes on time scales much shorter than the asymptotic
thermal equilibration time. For an anisotropic QCD plasma
Weibel instabilities may operate~\cite{Mrowczynski:1993qm}. 
For the reheating of the early
universe after inflation a tachyonic or parametric resonance
instability can lead to an exponential growth of occupation
numbers~\cite{Traschen:1990sw}.
Characteristic far-from-equilibrium phenomena, such as an
early prethermalization of the equation of state, have been
quantitatively studied in that context~\cite{Berges:2004ce,Podolsky:2005bw}.
This is followed by a comparably long quasi-stationary period in a manner
reminiscent of Kolmogorov wave turbulence~\cite{Micha:2002ey,Arnold:2005ef}.
The turbulent behavior is described by stationary solutions of classical
statistical field theory. Quantum corrections finally lead to deviations and
thermalization to Bose-Einstein or Fermi-Dirac 
distributions~\cite{Berges:2002cz,Arrizabalaga:2004iw,Berges:2000ur,Berges:2004ce}.

The presence of strong interactions and/or large occupation
numbers from nonequilibrium instabilities require nonperturbative
approximations. Aspects of systems with high occupation numbers
can be nonperturbatively described using classical field theory
methods. However, classical Rayleigh-Jeans divergences
and the lack of genuine quantum effects --- such as the approach
to quantum thermal equilibrium characterized by Bose-Einstein or
Fermi-Dirac statistics --- limit their use. Gauge-string duality
offers a novel possibility for studying dynamical properties of
certain strongly interacting gauge theories, which has led to
interesting results~\cite{Policastro:2001yc}.

Standard nonperturbative approximations such as based on
$1/N$ expansions of the {\em one-particle irreducible} (1PI)
effective action can be secular in time and do not provide a valid
description, similar to perturbation theory. Suitable resummation
techniques can be efficiently formulated using the {\em
two-particle irreducible} (2PI) effective action~\cite{CJT}. 
This has led to successful descriptions of far-from-equilibrium
dynamics and subsequent thermalization in scalar and fermionic 
quantum field theories in various 
dimensions~\cite{Berges:2000ur,Aarts:2001qa,Berges:2001fi,Berges:2002cz,Berges:2004ce,Arrizabalaga:2004iw,Berges:2004yj,Gasenzer:2005ze}. 
They have also been used to compute transport coefficients~\cite{Aarts:2003bk} 
and to determine the range of validity of transport or
semi-classical approaches~\cite{BergesTransport,GreinerNPA,Markus}. 

A nonperturbative description of nonequilibrium quantum fields 
can be based on the 2PI $1/N$ expansion beyond leading order (LO), which  
has been worked out in
detail~\cite{Berges:2001fi,Aarts:2002dj,Berges:2002cz,Berges:2004ce,Arrizabalaga:2004iw,Berges:2004yj,Gasenzer:2005ze}. 
So far, the results from the 2PI $1/N$
expansion to NLO are the only ones that bridge nonequilibrium
instabilities at early times and thermal equilibrium at late 
times.
Recently the 2PI $1/N$ expansion has been pushed forward towards
NNLO for a quantum anharmonic oscillator~\cite{Aarts:2006cv}. A series of
precision tests has been performed by 
now~\cite{Aarts:2001yn,Blagoev:2001ze,Arrizabalaga:2004iw},
including the computation of critical
exponents near second-order phase transitions~\cite{Alford:2004jj}. The
results exhibit some remarkable properties of the 2PI-$1/N$ expansion, curing
for instance the spurious small-$N$ divergence of the anomalous dimension seen
in the standard 1PI $1/N$ to NLO.  Recently, also the question of the
formation of topological defects has been investigated in the large-$N$
expansion for $N=1,2$ in one spatial dimension at zero temperature. Since
there are no defects in the corresponding quantum theory at zero temperature,
the authors drop quantum corrections and introduce a cooling procedure.
Defects should then appear as an artefact of the classical approximation,
however, the authors see no sign of them. The size of the observed
discrepancies are in accordance with earlier comparisons within classical
statistical field theory~\cite{Rajantie:2006gy}.
An important source of quantitative differences arises from the fact that a
large-$N$ expansion to NLO in the limit $N \to 1$ trivially misses $1/3$ of
the topologically equivalent diagrams starting at NNLO~\cite{Aarts:2002dj}. 

A nonperturbative large-$N$ expansion of the 2PI effective action 
beyond LO, feasible for the (abelian)
case of large numbers of flavors, seems not within reach for $SU(N)$ gauge
theories relevant for QCD. In this context it 
remains an important open question to determine
the range of validity of 2PI loop expansions, in particular, in view of a
number of remarkable results from related Dyson-Schwinger equations 
in thermal equilibrium~\cite{Fischer:2006ub}.
It has been an important step that
renormalization of 2PI resummed approximations for gauge theories
has been worked out~\cite{Reinosa:2006cm,Pawlowski:2005xe}, 
extending earlier work for
scalar~\cite{vanHees:2001ik} and fermionic theories~\cite{fermionrenorm}.

Complementary to approaches based on analytic approximations, one
of the outstanding problems in nonequilibrium quantum field theory
are first-principles simulations on a Minkowski space-time lattice. 
In the following we will describe recent efforts
along these lines, which are still in its infancies.

\section{Lattice simulations of real-time quantum fields}

Lattice gauge theory calculations are typically based on a 
Euclidean formulation,
where the time variable is analytically continued to imaginary
values. By this the quantum theory is mapped onto a statistical
mechanics problem, which can be simulated by importance sampling
techniques. Recovering real-time properties from the Euclidean
formulation is a formidable problem. 
Direct simulations in Minkowski space-time would be a
breakthrough in our efforts to resolve pressing questions, such as
early thermalization or the origin of seemingly perfect fluidity
in a QCD plasma at RHIC. For real times standard
importance sampling is not possible because of a non-positive
definite probability measure. Efforts to circumvent this problem
include mimicking the real-time dynamics by computer-time
evolution in Euclidean lattice
simulations~\cite{computertime,Miller:2000pd}. A problem in this
case is to calibrate the computer time independently of the
algorithm. In principle, direct simulations in Minkowski
space-time may be obtained using stochastic quantization
techniques, which are not based on a probability
interpretation~\cite{stochquant,Seiler:1983mz}.

\begin{figure}[t]
\vspace{4.9cm} \includegraphics{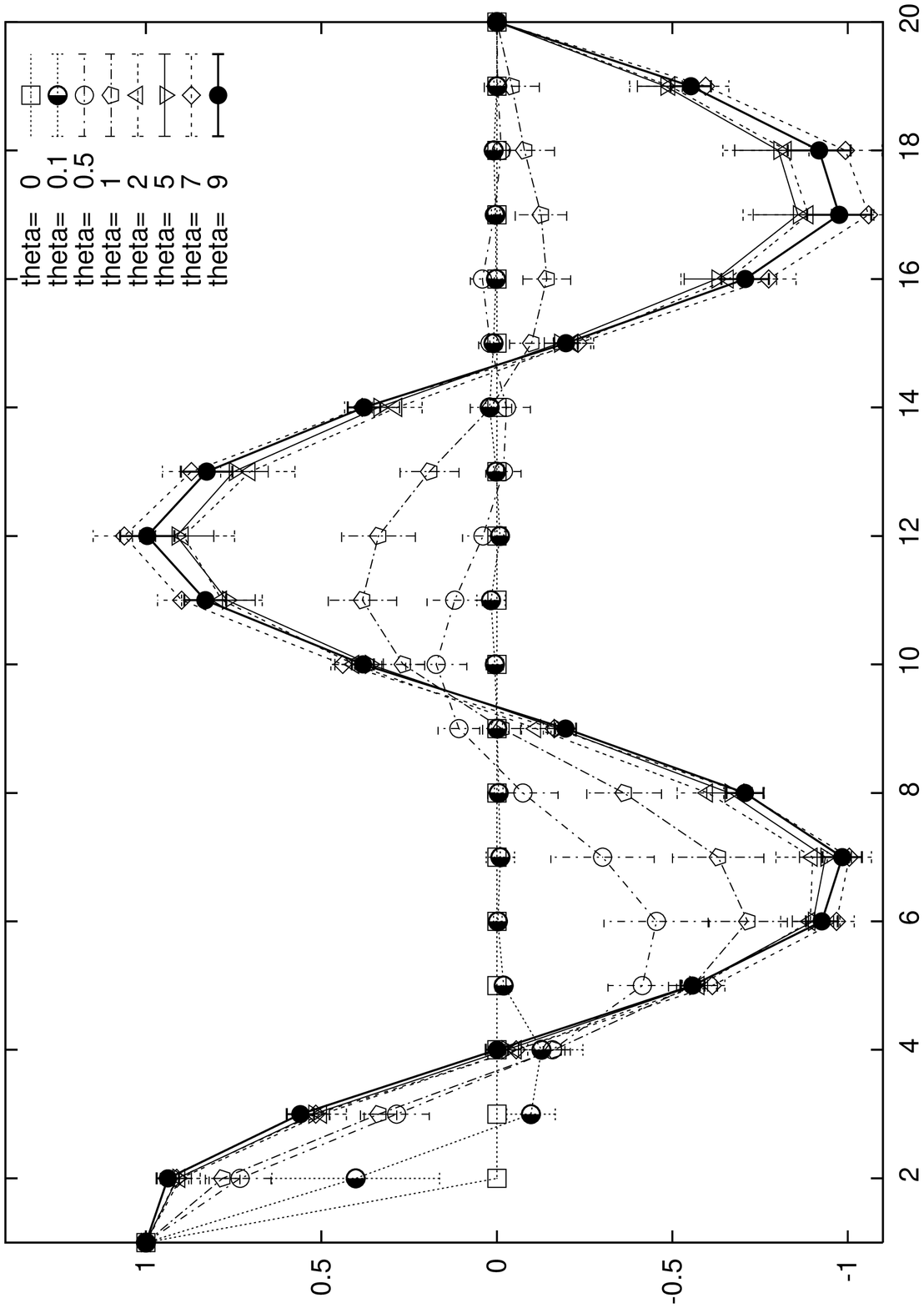}
\includegraphics{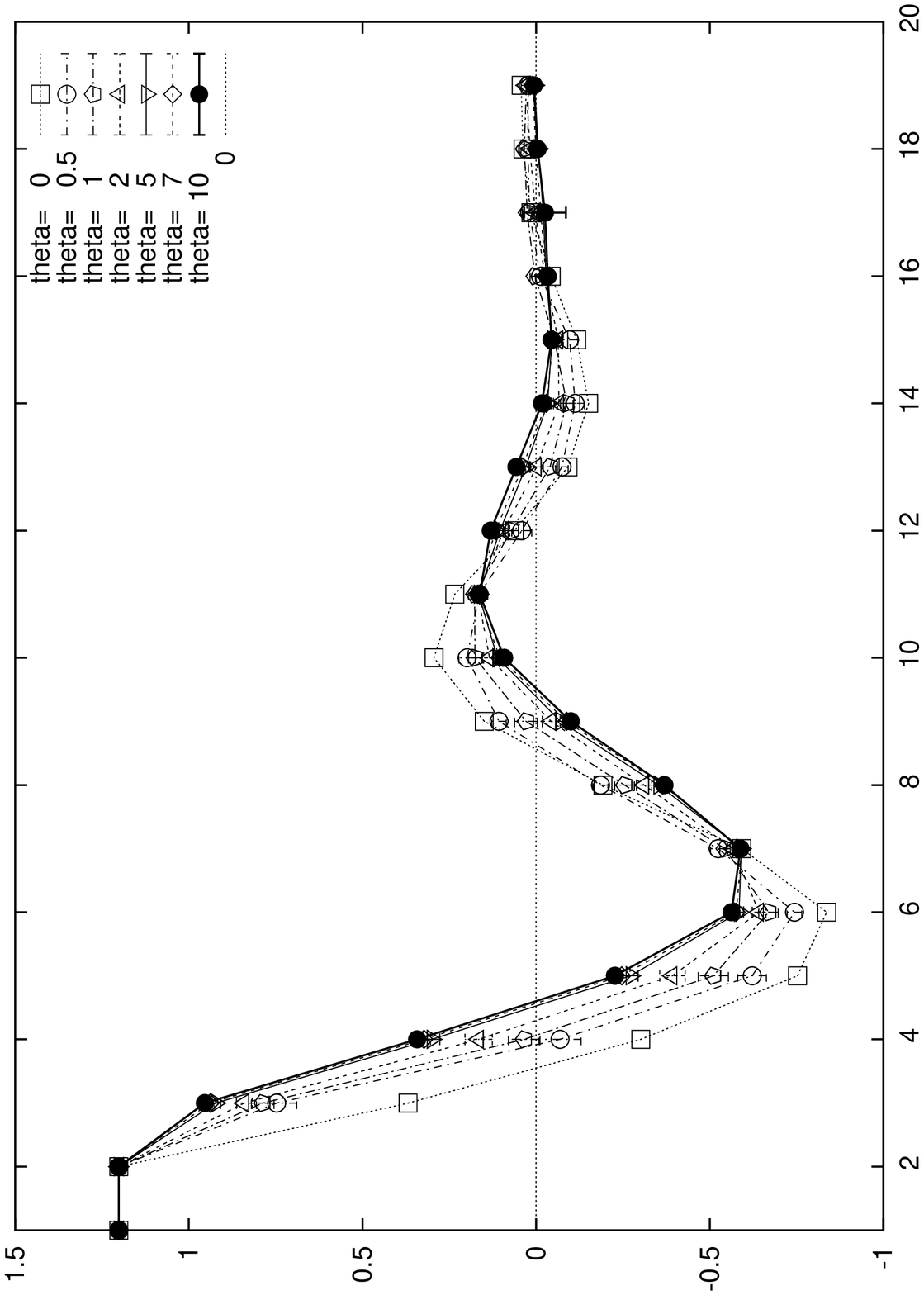} \caption{The real part
of $\langle \int_x \varphi(0,\bx) \varphi(t,\bx) \rangle$ as a funtion
of real time $t$ in units of the lattice spacing. Shown are snapshots at different Langevin times $\vartheta$. The LHS corresponds to free field theory, with null starting configuration at $\vartheta =0$. The RHS is for the interacting theory. Here the nonequilibrium classical statistical result is taken
as starting configuration, and the Langevin updating for $\vartheta > 0$ 
incorporates quantum corrections~\cite{Berges:2005yt}.} \label{fig:int}
\end{figure}

In Ref.~\cite{Berges:2005yt} this has been recently used to explore
nonequilibrium dynamics of an interacting scalar quantum field
theory. As an example, Fig.~\ref{fig:int} shows the time evolution
for the connected part of the unequal-time correlator ${\rm Re}
\langle \int_x \varphi(0,\bx) \varphi(t,\bx) \rangle$, which measures
the correlation of the field $\varphi$ at time $t$ with the initial
field. Shown are snapshots of the correlator at different  
Langevin-times. The left figure shows the result for the free field theory with null start configuration. The Langevin updating reproduces the correct
free-field result. The right figure gives an example for the interacting 
theory, which exhibits a finite characteristic damping time.
One observes good convergence properties of the quantum simulations,
which is a remarkable result. For given initial field
configurations at time $t=0$, very different starting
configurations for the ($3+1$)-dimensional space-time lattice converge
to the same nonequilibrium dynamics for all $t > 0$~\cite{Berges:2005yt}. 

\subsection{Real-time stochastic quantization}

In real-time stochastic quantization the quantum ensemble is
constructed by a stochastic process in an additional
``Langevin-time'' using the reformulation for the Minkowskian path
integral~\cite{cl,Minkowski}: The quantum fields are defined on a
$d$-dimensional physical space-time lattice, while the updating
procedure employs a Langevin equation with a complex driving force
in an additional, unphysical ``time'' direction. Though more or
less formal proofs of equivalence of the stochastic approach and
the path integral formulation have been given for Minkowski
space-time, not much is known about the general convergence
properties and its reliability beyond free-field theory or simple
toy models~\cite{Minkowski}. More advanced applications concern
simulations in Euclidean space-time with non-real
actions~\cite{Ambjorn:1986mf,Karsch:1985cb}.

Below we follow Ref.~\cite{newpaper} and 
discuss real-time stochastic quantization for a
scalar theory and $SU(N)$ pure gauge theory relevant for
QCD. For a scalar theory with Minkowski action $S[\varphi]$ the
Langevin updating equation reads
\begin{equation}
\varphi'(x) = \varphi(x) + i\, \epsilon \, \frac{\delta
S[\varphi]}{\delta \varphi(x)} + \sqrt{\epsilon}\, \eta(x) \,
, \label{eq:complexlange}
\end{equation}
with Gaussian noise 
\beq 
\langle \eta(x)
\rangle_\eta = 0 \,, \qquad \langle \eta(x)\, \eta(y) \rangle_\eta
= 2\, \delta(x-y) \, . \label{eq:noise} 
\eeq
The sum over all Langevin steps, $\vartheta \equiv \sum
\epsilon$, corresponds to Langevin-time.

For $SU(N)$ gauge theory on a $(N_s a_s)^3 \times N_t a_t$ lattice the
real-time classical action reads
\begin{eqnarray}
S[U] &=& - \beta_0 \sum_{x} \sum_i \left\{ \frac{1}{2 {\rm Tr}
{\bf 1}} \left( {\rm Tr}\, U_{x,0i} + {\rm Tr}\, U_{x,0i}^{-1}
\right) - 1 \right\}
\nonumber\\
&& + \beta_s \sum_{x} \sum_{i,j \atop i<j} \left\{ \frac{1}{2 {\rm
Tr} {\bf 1}} \left( {\rm Tr}\, U_{x,ij} + {\rm Tr}\, U_{x,ij}^{-1}
\right) - 1 \right\} \, ,
\end{eqnarray}
with spacial indices $i,j = 1,2,3$. It is described in terms of
the gauge invariant plaquette variable
\begin{equation}
U_{x,\mu\nu} \equiv U_{x,\mu} U_{x+\hat\mu,\nu}
U^{-1}_{x+\hat\nu,\mu} U^{-1}_{x,\nu} \label{eq:plaq}\, .
\end{equation}
Here $U_{x,\mu}$ is the
parallel transporter associated with the link from the
neighbouring lattice point $x+\hat{\mu}$ to the point $x$ in the
direction of the lattice axis $\mu = 0,1,2,3$. The couplings are
\begin{equation}
\beta_0 \equiv \frac{2 \gamma {\rm Tr} {\bf 1}}{g_0^2} \,\, ,
\quad \beta_s \equiv \frac{2 {\rm Tr} {\bf 1}}{g_s^2 \gamma} \, ,
\label{eq:ganiso}
\end{equation}
with the anisotropy parameter $\gamma \equiv a_s/a_t$. 
The Langevin updating equation for $U_{x, \mu}$ then reads
\begin{eqnarray}
U^{\prime}_{x, \mu} &=& \exp\left\{ i \lambda_a \left(\epsilon\, i
D_{x \mu a} S[U] + \sqrt{\epsilon}\, \eta_{x \mu a}
\right)\right\} U_{x, \mu}\, , \label{eq:ALangevin}
\end{eqnarray}
with
\begin{eqnarray}
i D_{x \mu a} S[U] &=& - \frac{1}{2N} \sum_{\nu=0 \atop \nu \neq \mu}^{3}
\beta_{\mu\nu} {\rm Tr} \left( \lambda_a U_{x,\mu} C_{x,\mu\nu}-
\bar{C}_{x,\mu\nu} U^{-1}_{x,\mu} \lambda_a \right) \, .
\label{eq:derivS}
\end{eqnarray}
For a compact notation we have defined $\beta_{ij} \equiv \beta_s$
and $\beta_{0i} \equiv \beta_{i0} \equiv - \beta_0$ and
\begin{eqnarray}
C_{x,\mu\nu} &=& U_{x+\hat\mu,\nu} U^{-1}_{x+\hat\nu,\mu}
U^{-1}_{x,\nu} + U^{-1}_{x+\hat\mu-\hat\nu,\nu}
U^{-1}_{x-\hat\nu,\mu} U_{x-\hat\nu,\nu}
\nonumber\\
\bar{C}_{x,\mu\nu} &=&
U_{x,\nu}U_{x+\hat\nu,\mu}U^{-1}_{x+\hat\mu,\nu} +
U^{-1}_{x-\hat\nu,\nu} U_{x-\hat\nu,\mu}U_{x+\hat\mu-\hat\nu,\nu}
\, .
\end{eqnarray}

It is shown in Ref.~\cite{newpaper} (see also Ref.~\cite{xue}) that stationary 
solutions of equations (\ref{eq:complexlange}) or (\ref{eq:ALangevin})
always fulfill the infinite set of (symmetrized) Dyson-Schwinger 
identities of the respective quantum field theory. For instance, 
for the scalar theory one finds
from the fixed points of the Langevin equation (\ref{eq:complexlange}) the
infinite hierarchy of identities
\bea 
\left\langle \frac{\delta
S[\varphi]}{\delta \varphi(x)}\right\rangle_\eta &=& 0 \,,
\label{eq:sd1} 
\eea 
\bea
\left\langle \frac{\delta S[\varphi]}{\delta
\varphi(x)}\varphi(y) \right\rangle_\eta + \left\langle
\frac{\delta S[\varphi]}{\delta \varphi(y)}\varphi(x)
\right\rangle_\eta &=& 2 i \delta(x-y) \, , \label{eq:sd2}
\eea
\bea \left\langle\frac{\delta
S[\varphi]}{\delta \varphi(x)}\varphi(y)\varphi(z)
\right\rangle_\eta +\left\langle \frac{\delta
S[\varphi]}{\delta \varphi(y)}\varphi(x)\varphi(z)
\right\rangle_\eta +\left\langle \frac{\delta
S[\varphi]}{\delta
\varphi(z)}\varphi(x)\varphi(y) \right\rangle_\eta =&& \nonumber\\
2 i \left(\left\langle \varphi(x) \right\rangle_\eta \delta(y-z) +
\left\langle \varphi(y) \right\rangle_\eta \delta(x-z) +
\left\langle \varphi(z) \right\rangle_\eta \delta(x-y)\right) \, ,
\label{eq:sd3} \eea and correspondingly for the higher $n$-point
functions in Minkowski space-time. If the Langevin updating converges,
which is the case for the implementation of Ref.~\cite{newpaper}, then it 
solves the correct infinite set of Dyson-Schwinger identities. 
In addition, {\em Euclidean} theories with real
actions can be shown to have a unique solution based on positivity
arguments. A similar argument fails for real-time
stochastic quantization. In general, here the correct fixed point
cannot be chosen a priori without implementing further
constraints.

\subsection{Precision tests for scalar and nonabelian gauge theories}

In Ref.~\cite{newpaper} tests for real-time stochastic quantization 
are discussed. 
Similar to what has been observed in Ref.~\cite{Berges:2005yt} for scalar
fields, also for pure gauge theory one finds that previously reported
unstable dynamics represents no problem in practice: A combination
of sufficiently small Langevin step size and the use of a "tilted"
real-time contour leads to converging results in general. The employed
procedure respects gauge invariance and appears to be well under
control. This is exemplified for $SU(2)$ gauge theory in $3+1$
dimensions and for a scalar theory in zero spatial dimension,
i.e.\ a quantum anharmonic oscillator. For the latter the 
Schr{\"o}dinger equation can be solved as well
numerically by diagonalization of the Hamiltonian, for comparison.
\begin{figure}[t]
\begin{center}
\epsfig{file=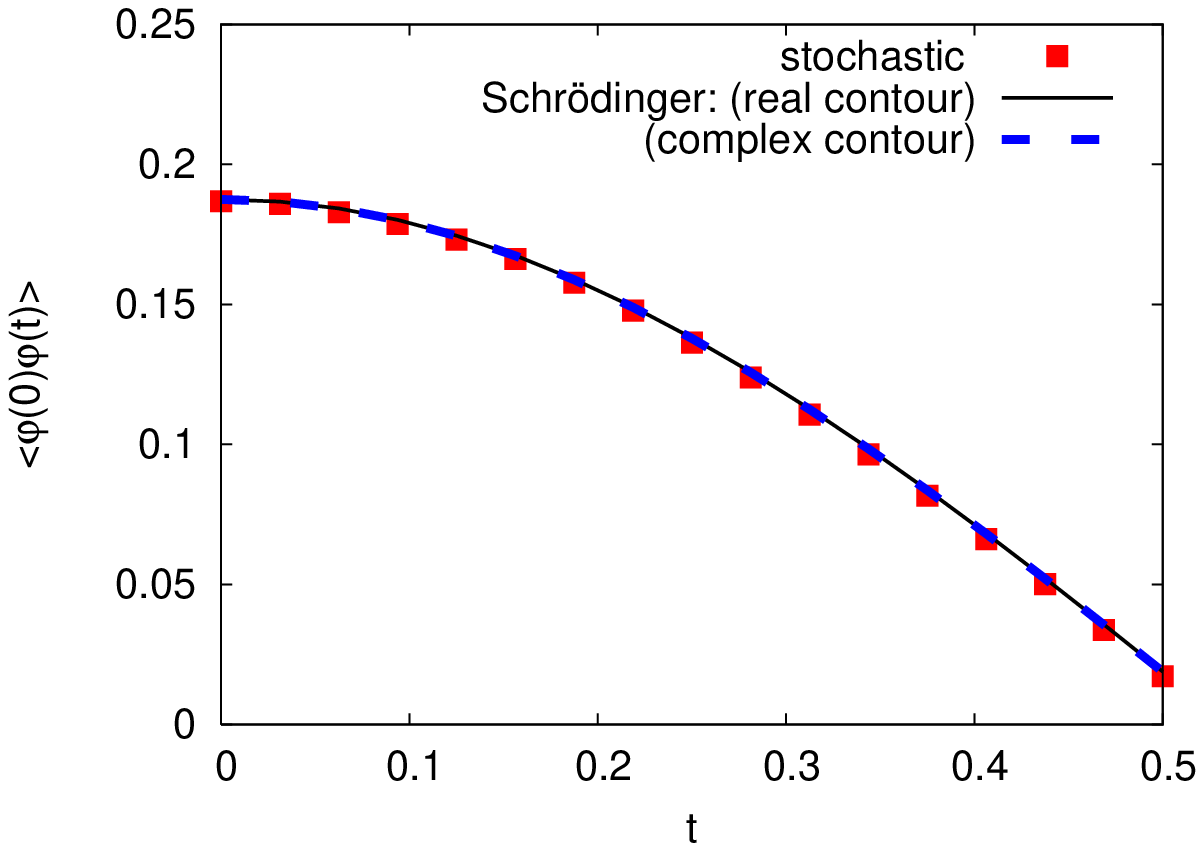,width=7.9cm}
\epsfig{file=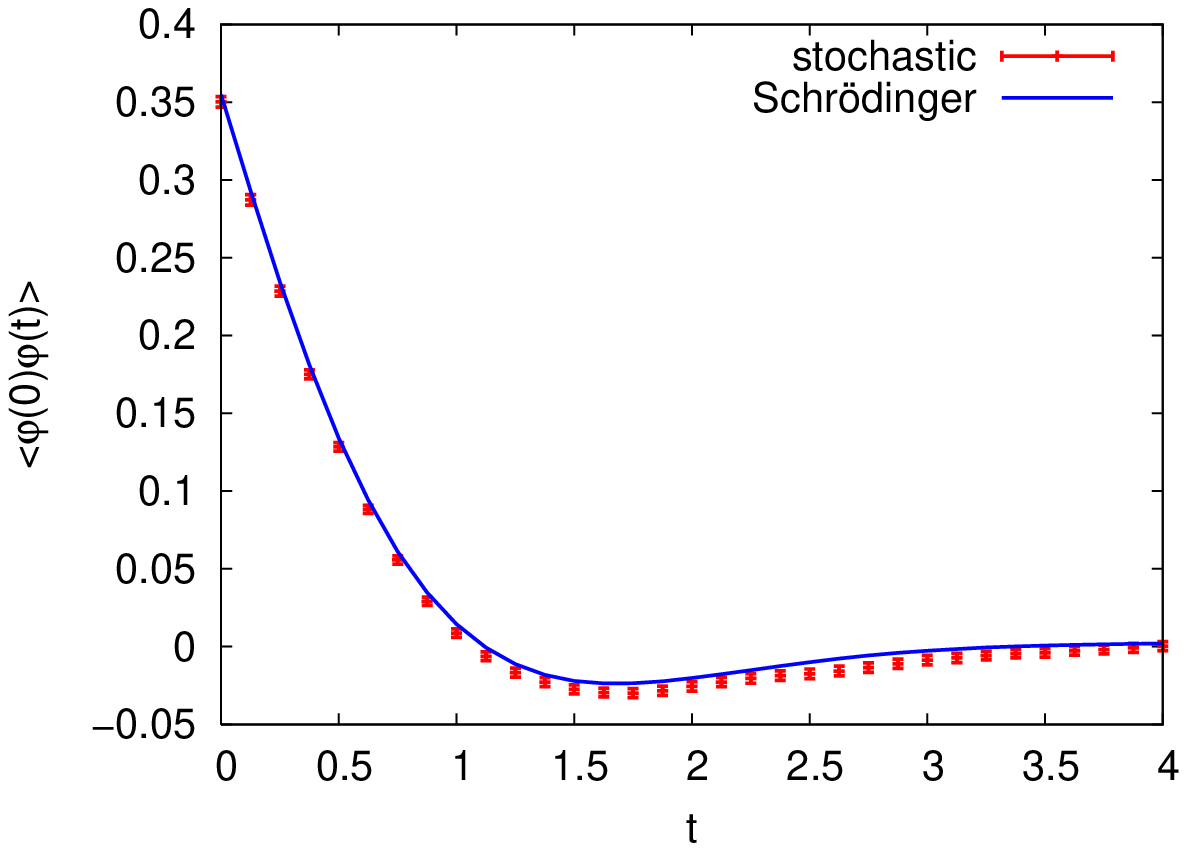,width=7.9cm}
\vspace*{-1cm}
\end{center}
\caption{The unequal-time two-point
correlation function for the quantum anharmonic oscillator as a
function of real time $t$. The (short-time) results obtained from stochastic
quantization agree to very good accuracy to those obtained from
directly solving the Schr{\"o}dinger equation. For the left figure  
a strong coupling $\lambda = 96$ and $\beta = 1$ is employed, whereas 
$\lambda = 6$ and $\beta = 8$
for the right.~\cite{newpaper}} \label{fig:T4dh}
\end{figure}

For the scalar theory one finds that stochastic quantization
accurately describes the time evolution for lattices with
sufficiently small real-time extent. This concerns nonequilibrium
and equilibrium simulations at weak as well as strong couplings.
As an example, Fig.~\ref{fig:T4dh} shows results for the time evolution for 
$\lambda \varphi^4/4!$ scalar theory in zero spatial dimension. Shown is  
the two-point correlation function
$\langle \varphi(0) \varphi(t) \rangle_\eta$ as a function of real
time $t$. The simulations are done in thermal equilibrium, 
in which case the real-time contour extends along the imaginary axis to the
inverse temperature $\beta$. For the left figure the real-time extent
of the contour is $\Delta t = 0.5$. Here the upper branch of the
contour has a tilt of $0.001 \beta$, so it is almost horizontal
and, therefore, realizes to high accuracy a real-time contour. For
comparison to the stochastic quantization result the
time evolution from the Schr{\"o}dinger equation in Minkowski time
as well as using the employed complex times from the corresponding
matrix algebra is given. They all agree to very good accuracy.

However, when the real-time extent of the lattice is enlarged the
stochastic updating does not converge to the correct solution. 
Fig.~\ref{fig:T4dh} shows the two-point correlation function
as a function of real time $t$ for $\lambda = 24$ and $\beta = 1$.
For a short real-time lattice corresponding to $t_{\rm final}= 0.8$
one observes from the left figure excellent agreement of stochastic
quantization and Schr{\"o}dinger equation results. However, doubling the  
extent of the contour leads to a qualitatively different behavior as
given in the right plot of Fig.~\ref{fig:T4dh}. This difference persists 
also on finer real-time grids. The non-vanishing 
imaginary part of the equal-time correlator and the loss of time-translation 
invariance reflects a non-unitary time evolution.
\begin{figure}[t]
\begin{center}
\epsfig{file=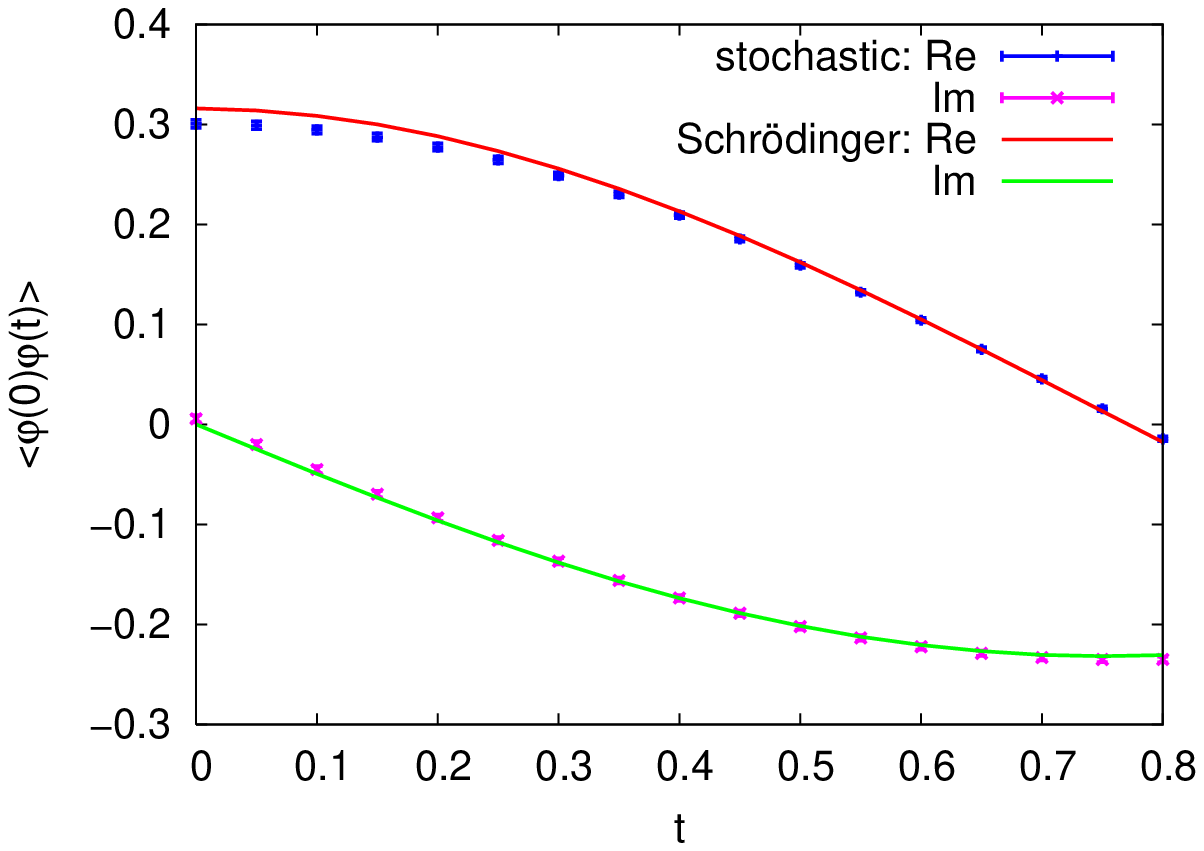,width=7.9cm}
\epsfig{file=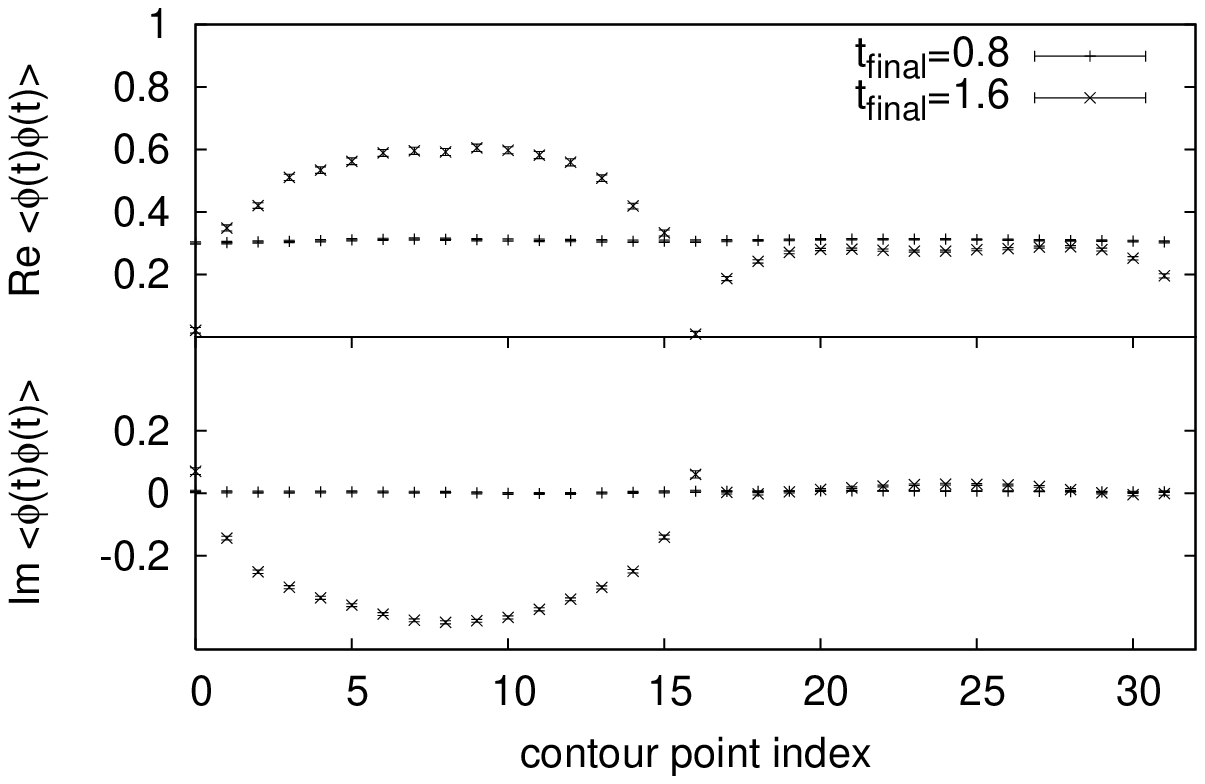,width=7.9cm}
\vspace*{-1cm}
\end{center}
\caption{Real and imaginary part of the two-point
correlation function for the quantum anharmonic oscillator as a
function of real time $t$. Left: for short real-time extent of the lattice the
stochastic quantization results agree well to those obtained from
the Schr{\"o}dinger equation. Right: Equal-time 
correlator 
as a function of time (index of lattice site along the time-contour). 
Compared are two simulations where the real-time extent of the lattice 
differs by a factor of two. The larger lattice leads to a qualitatively 
different, non-unitary behavior.~\cite{newpaper}} \label{fig:mean_T4dhp}
\end{figure}

\begin{figure}[t]
\begin{center}
\epsfig{file=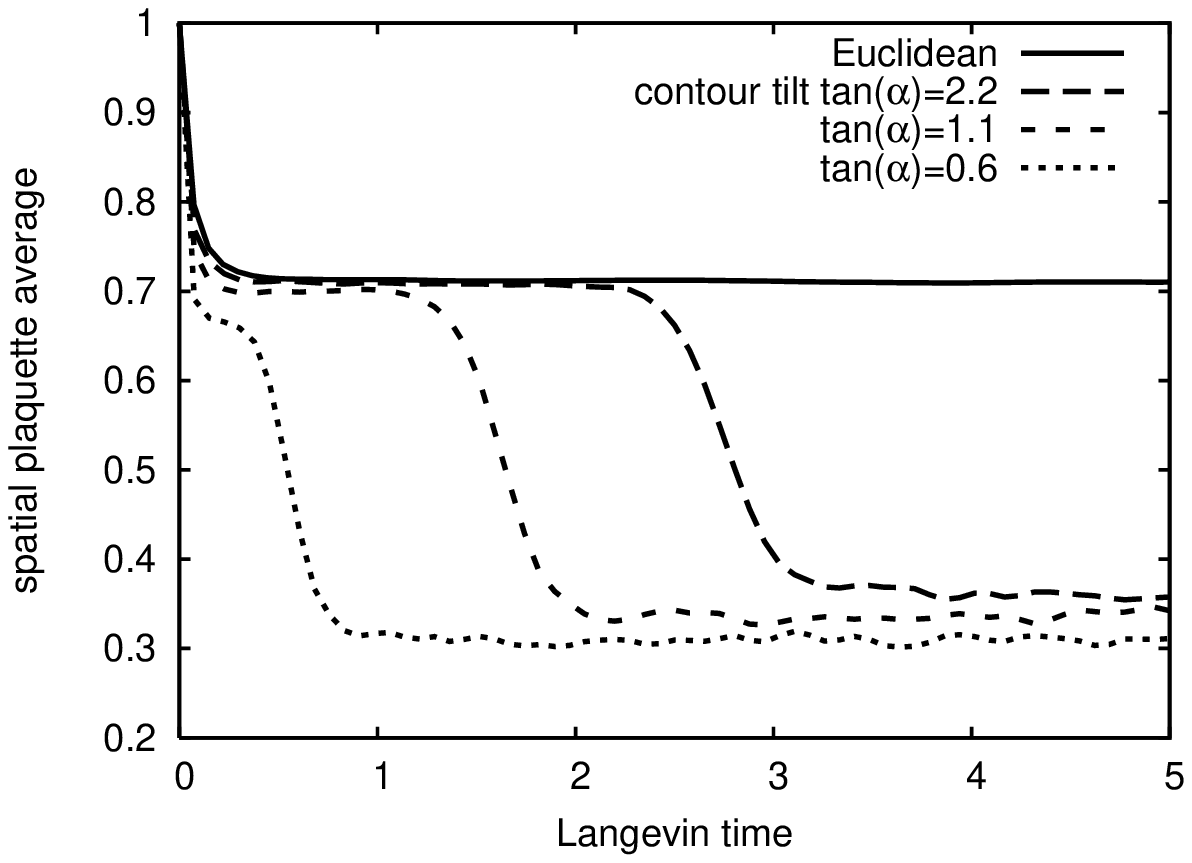,width=7.9cm}
\epsfig{file=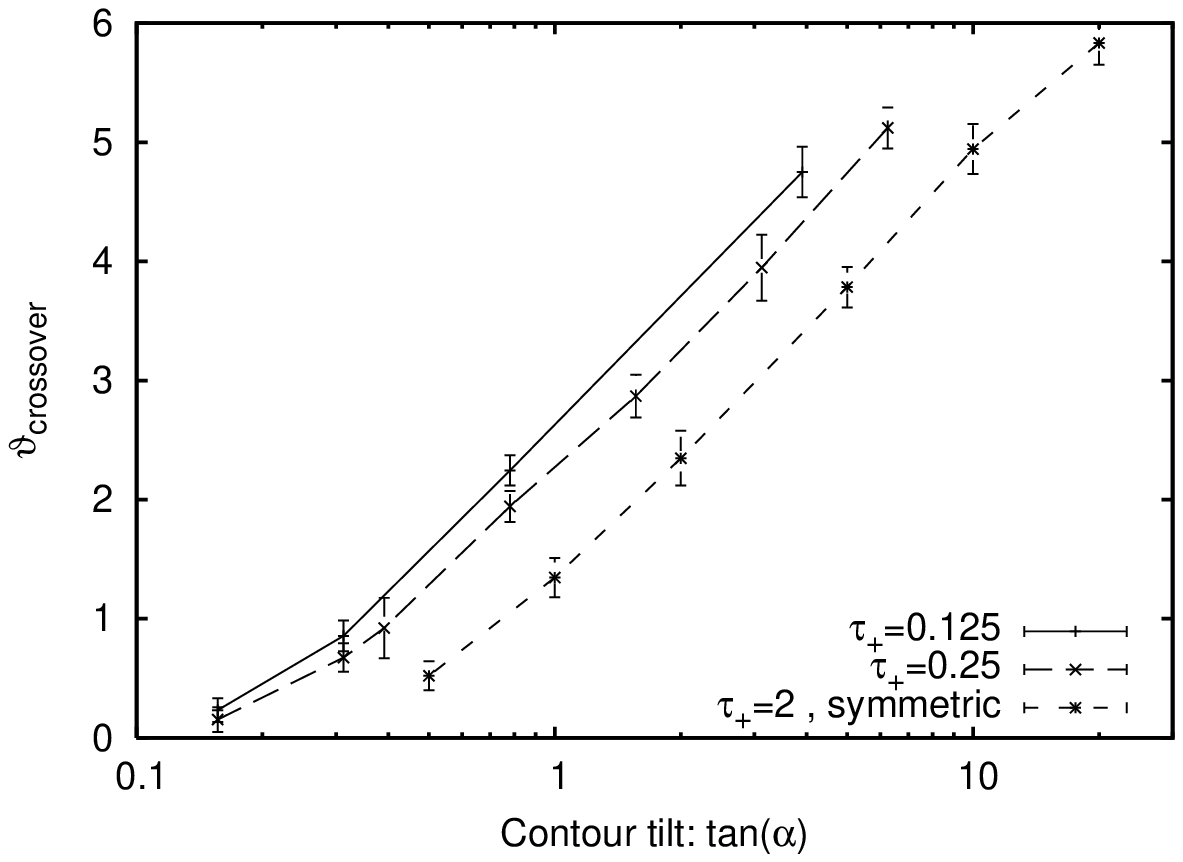,width=7.9cm}
\vspace*{-1cm}
\end{center}
\caption{The spatial plaquette average for $SU(2)$ gauge theory 
in thermal equilibrium as a function of 
Langevin time. Shown are results for different complex contours. 
Here $\alpha = 0$ corresponds to a contour with an infinite extent 
along the real-time axis, while $\alpha = \pi/2$ denotes the Euclidean 
contour. The longer the real-time component of the contour the less 
accurate the thermal solution is approached. Here $\vartheta_{\rm crossover}$
denotes the Langevin-time at which
the crossover to the 
non-unitary fixed point occurs.~\cite{newpaper}} \label{fig:2fixp}
\end{figure}

Doing the equivalent investigation for $SU(2)$ pure gauge theory for 
thermal equilibrium, one finds a similar behavior with an important 
difference. The left of Fig.~\ref{fig:2fixp} shows the 
{\em Langevin-time} evolution of the spatial plaquette average. This quantity 
is computed for different complex contours, with the approximation that
$g_0 = g_s$. The solid line shows the 
result for vanishing real-time extent, i.e.\ for a Euclidean contour 
corresponding to an inverse temperature $\beta = 4$. 
The different dashed curves correspond to results for complex contours on 
isosceles triangles each having a different tilt $\alpha$ with respect 
to the real-time axis. Here $\alpha = 0$ would correspond to an infinite 
extent along the real axis. One observes that with increased real-time 
extent or smaller tilt $\alpha$ the correct thermal solution is 
approached less accurately. In particular, it is only approached at 
intermediate Langevin-times, irrespective of the details of the non-vanishing
real-part of the contour. This aspect differs from the scalar case where short
real-time extents lead to stable thermal solutions.

\begin{figure}[t]
\begin{center}
\epsfig{file=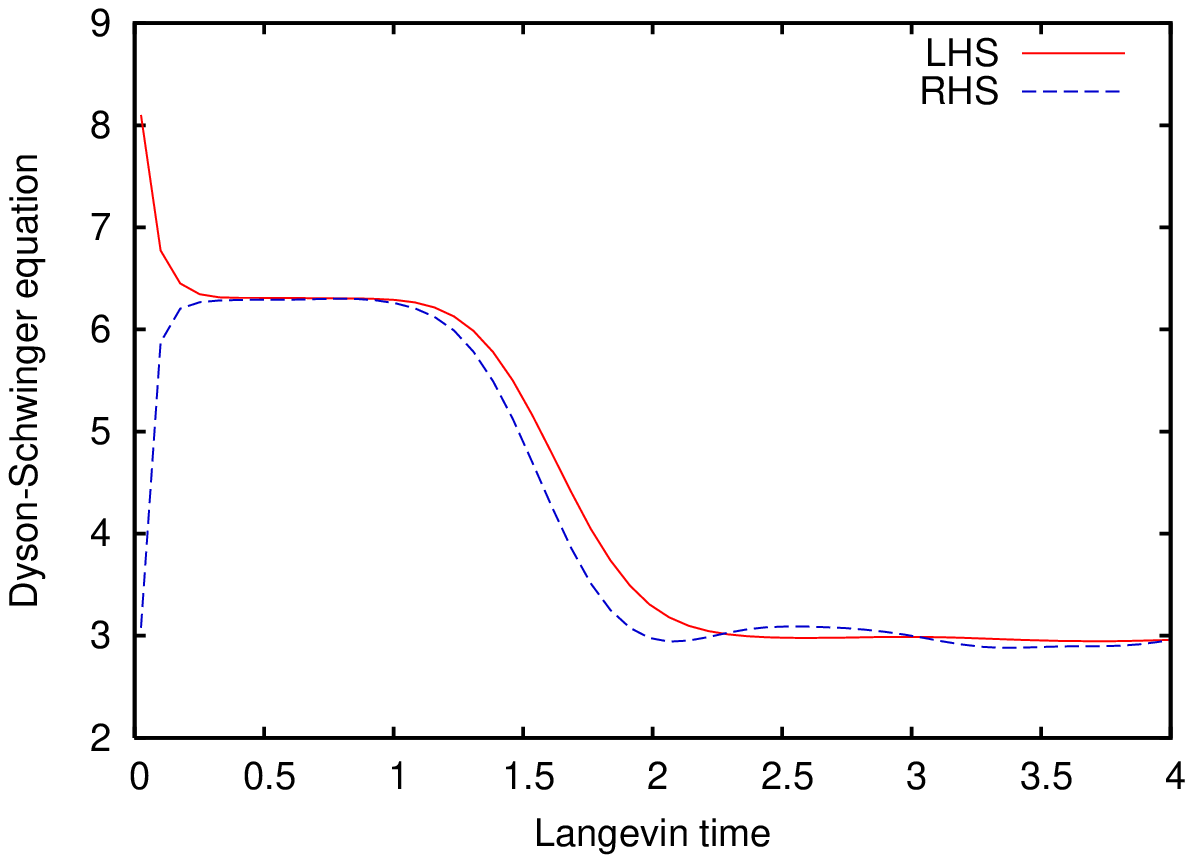,width=7.9cm}
\epsfig{file=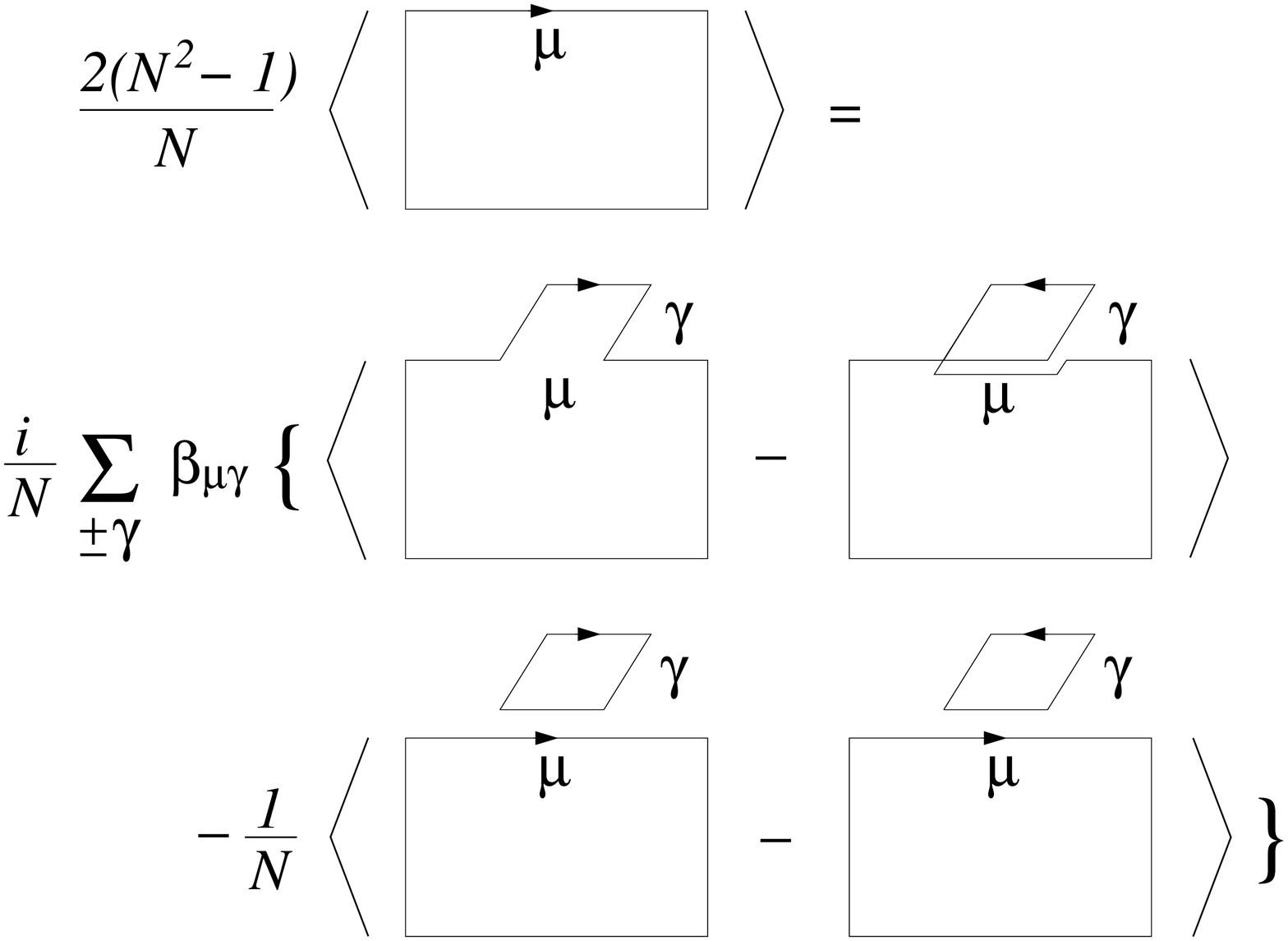,width=7.cm}
\vspace*{-1cm}
\end{center}
\caption{Numerical measurement (left) of the Dyson-Schwinger 
equation (right) for a spatial plaquette 
variable.~\cite{newpaper}} \label{fig:DSnum}
\end{figure}

As mentioned above, all converging solutions of the 
stochastic dynamics method fulfill the same infinite set of
(symmetrized) Dyson-Schwinger identities of the quantum field
theory. This is remarkable in view of the different ``physical'' and
``unphysical'' solutions that are observed. In Fig.~\ref{fig:DSnum} 
this is visualized for the example of the 
Dyson-Schwinger equation for a spatial plaquette variable in $SU(2)$ 
gauge theory. Plotted are separately the LHS and the RHS of the 
Dyson-Schwinger equation displayed graphically on the right of 
Fig.~\ref{fig:DSnum}. The left plot displays the respective LHS and 
RHS as a function of Langevin-time for a typical contour with 
tilt $0 < \alpha < \pi/2$. The flow with Langevin-time 
quickly leads to a rather accurate agreement of both sides such that
the Dyson-Schwinger equation is fulfilled. However, after some 
Langevin-time they start deviating again, finally leading to another
stationary value where the LHS and RHS agree to reasonable 
accuracy. 

The evolution with Langevin-time is governed by fixed points 
and, for the typical case considered in Fig.~\ref{fig:DSnum}, 
the correct thermal 
fixed point is approached at first. However, it is not stable and the
Langevin flow exhibits a crossover to another (stable) fixed point. 

In conclusion, short real-time physics in thermal equilibrium can be 
reproduced reasonably well if the length of the real time contour is small on
the scale of the inverse temperature $\beta$. For longer contours the boundary
conditions in physical time do not seem to constrain enough the Langevin flow
and the 'life-time' of the thermal (physical) fixed point decreases 
(nonabelian gauge theory),
or the fixed point becomes fully unstable (scalar theory).
A second, apparently stable fixed point develops for large
Langevin times. This latter represents an
unphysical, non-unitary regime, to be recognized by non-translational 
invariant expectation functions and violation of 
the unsymmetrized Dyson-Schwinger identities (while the symmetrized ones
are still satisfied, indicating convergence)~\cite{newpaper}.
Further work has to concentrate on further means for
controlling and optimizing the method. The method allows for 
quite some flexibility as to which quantities
are chosen to define the stochastic process, or introducing
a stochastic re-weighting. 

We are indepted to D. Sexty and I-O. Stamatescu for 
collaboration on this topic.

\end{document}